\documentclass[11pt,twoside,showpacs,superscriptaddress,showkeys]{revtex4}
\usepackage{graphicx,latexsym,amssymb,amsmath,color, multirow, mathrsfs, booktabs}

\usepackage[section]{placeins}

\newcommand{\bpm}{\begin{pmatrix}}
\newcommand{\epm}{\end{pmatrix}}
\newcommand{\beq}{\begin{equation}}
\newcommand{\eeq}{\end{equation}}
\newcommand{\bea}{\begin{array}}
\newcommand{\eea}{\end{array}}
\newcommand{\bsub}{\begin{subequations}}
\newcommand{\esub}{\end{subequations}}

\newcommand{\cals}[1]{{\mathcal #1}}
\newcommand{\scr}[1]{{\mathscr #1}}
\newcommand{\ff}[1]{\frac{1}{#1}}

\newcommand{\lrb}[1]{\left(#1\right)}
\newcommand{\lrs}[1]{\left[#1\right]}

\newcommand{\svec}[1]{{\mbox{\boldmath${ #1}$}}}

\newcommand{\re}{\nonumber\\}

\newcommand{\tabref}[1]{{\sf\bfseries Table \ref{#1}}}
\newcommand{\figref}[1]{{\sf\bfseries Fig. \ref{#1}}}

\newcommand{\pcb}{{\text{PCB}}}
\newcommand{\pso}{{\text{PSO}}}
\newcommand{\so}{{\text{SO}}}

\begin{document}

\title{Pseudo-spin symmetry in density-dependent relativistic Hartree-Fock theory}
\author{WenHui Long}\email{whlong@pku.org.cn}
\affiliation{School of Physics, Peking University, 100871 Beijing, China} \affiliation{Center for
Mathematical Sciences, University of Aizu, Aizu-Wakamatsu, 965-8580 Fukushima, Japan}
\affiliation{Institut de Physique Nucl$\acute{e}$aire, CNRS-IN2P3, Universit$\acute{e}$ Paris-Sud,
91406 Orsay, France}
\author{Hiroyuki Sagawa}
\affiliation{Center for Mathematical Sciences, University of Aizu, Aizu-Wakamatsu, 965-8580
Fukushima, Japan}
\author{Jie Meng}
\affiliation{School of Physics, Peking University, 100871 Beijing,
China}
 \affiliation{Institute of Theoretical Physics, Chinese Academy of Sciences, Beijing, China}
 \affiliation{Center of Theoretical Nuclear Physics, National Laboratory of Heavy Ion Accelerator, 730000 Lanzhou, China}
\author{Nguyen Van Giai}
\affiliation{Institut de Physique Nucl$\acute{e}$aire, CNRS-IN2P3, Universit$\acute{e}$ Paris-Sud,
91406 Orsay, France}

\begin{abstract}
The pseudo-spin symmetry (PSS) is investigated in the density-dependent
relativistic Hartree-Fock theory by taking {the} doubly magic nucleus
$^{132}$Sn as a representative. It is found that the Fock terms bring
significant contributions to the pseudo-spin orbital potentials (PSOP) and make
it comparable to the pseudo-centrifugal barrier (PCB). However, these Fock
terms in the PSOP are counteracted by other exchange terms due to the
non-locality of the exchange potentials. The pseudo-spin orbital splitting
indicates that the PSS is preserved well for the partner states $\lrb{\nu
3s_{1/2}, \nu2d_{3/2}}$ of $^{132}$Sn in the relativistic Hartree-Fock theory.

\end{abstract}
 \maketitle


The pseudo-spin symmetry (PSS), quasi-degeneracy of two single
particle states with the quantum numbers $(n, l, j = l + 1/2)$ and
{$({\bar n}=n-1,{\bar l}= l + 2, {\bar j}= j = l + 3/2)$}, was
firstly discovered more than thirty years ago in spherical nuclei
\cite{Arima:1969, Hecht:1969} {and later in deformed nuclei
\cite{Bohr:1982,Beuschel:1997}}. During the past ten years, {many}
efforts have been {made to investigate} the origin of PSS
\cite{Ginocchio:1997, Ginocchio:1998, Ginocchio:1998prc,
Ginocchio:1999} and the conservation conditions
\cite{Ginocchio:1997, Meng:1998PRCps, Tanabe:1998, Meng99,
Marcos:2001} within the framework of relativistic mean field (RMF)
theory. In Ref. \cite{Ginocchio:1997}, the PSS was interpreted as
the relativistic symmetry in the Dirac equation, which arises from
the cancellation between an attractive scalar potential $\Sigma_S$
and a repulsive vector {potential} $\Sigma_0$, i.e., $\Sigma_S +
\Sigma_0 {\simeq} 0$, and the pseudo-orbit number $\tilde l$ is
nothing but the orbital angular momentum of the lower component of
the Dirac wave function. However, in this limit there {is} no bound
states in the mean field, {at variance with} reality. More realistic
conditions were discussed for the conservation of the PSS in Refs.
\cite{Meng:1998PRCps, Tanabe:1998, Marcos:2001}.

In the framework of the RMF theory \cite{Miller:1972, Walecka:1974, Serot:1986,
Reinhard:1989, Ring:1996, Serot:1997, Bender:2003, Meng:2006}, i.e., the
relativistic Hartree approach with a no-sea approximation, the nucleons
interact via the exchange of mesons and photons. For the description of nuclear
structure, the relevance of relativity is not {in the need of} relativistic
kinematics but {it lies in} a covariant formulation which maintains the
distinction between scalar and vector fields (more precisely, the zeroth
component of {the Lorentz four-vector field}). The representations with large
scalar and vector fields in nuclei {(of the order of a few hundred MeV)}
provide more efficient descriptions of nuclear systems than non-relativistic
approaches, for example the origin of the nuclear spin-orbit potential
\cite{Lalazissis:1998} and that of the PSS \cite{Ginocchio:1997,
Meng:1998PRCps}.

Although there exist some attempts to include the exchange terms in the
relativistic description of nuclear matter and finite nuclei
\cite{Bouyssy:1987, Bernardos:1993,Marcos:2004}, the relativistic Hartree-Fock
(RHF) method was still not comparable with the RMF theory in the quantitative
description of nuclear systems. Recently, it was shown that the
density-dependent relativistic Hartree-Fock (DDRHF) theory \cite{long:2006a}
gives a successful quantitative description of nuclear matter and finite nuclei
at the same level as the RMF \cite{long:2006b}. Compared with RMF, the
relevance of the relativity is still kept well in DDRHF although the covariant
formulation becomes much more complicated due to the exchange terms. Since the
DDRHF {describes quite well}
nuclear systems, it is worthwhile to investigate the role of exchange terms in
the PSS, especially the influence of the non-locality on the conservation of
the PSS. In this {work}, we study the role of the Fock terms on the
conservation of PSS in the covariant relativistic Hartree-Fock theory. {The
numerical study is done with the effective Lagrangian PKO1 \cite{long:2006a} in
the case of DDRHF, and the results are compared with those obtained with the
RMF model PKDD \cite{Meng:2006}.}

For spherical nuclei, the Dirac spinor can be written as,
 \beq\label{SpinorSP}
f_\alpha(\svec r) =\ff r \bpm iG_a (r) \scr Y_{j_am_a}^{l_a}(\hat{\svec r})\\
-F_a(r)\scr Y_{j_am_a}^{l_a'}(\hat{\svec r})\epm \chi_{\ff2}( \tau_a),
 \eeq
where $\chi_{\ff2}(\tau_a)$ is the isospinor, $G_a$ and $F_a$ correspond to the
radial parts of upper and lower components, respectively. $\scr
Y_{j_am_a}^{l_a}$ is the spinor spherical harmonics and $\scr
Y_{j_am_a}^{l_a'}(\hat{\svec r})= -\hat{\svec\sigma} \centerdot \hat{\svec
r}\scr Y_{j_am_a}^{l_a}(\hat{\svec r})$ with $l_a' = 2j_a - l_a$. In the
spherical nuclei, the total angular momentum $j_a$, its projection on the $z$
axis $m_a$ and $\hat\kappa = -\hat\beta(\svec{\hat \sigma}\cdot\hat{\svec
L}+1)$ are conserved. The eigenvalues of $\hat \kappa$ are $\kappa_a = \pm
(j_a+\ff2)$ ($-$ for $j_a=l_a+\ff2$ and $+$ for $j_a = l_a-\ff2$). In the
following, the sub-index $a$ will be omitted {in the notations.}

Within the DDRHF, the radial Dirac equations, i.e., the relativistic
Hartree-Fock equations for spherical nuclei, are {expressed} as the coupled
differential-integral equations \cite{Bouyssy:1987, long:2006b}
 \bsub\label{RHF1}\begin{align}
E G(r)  =& -\lrs{\frac{d}{dr}-\frac{\kappa }{r}} F(r) \re & + \lrs{
\Sigma_{S}(r) + \Sigma_{0}(r)} G (r) + Y(r) \label{HF1},\\
E F(r)  =& +\lrs{\frac{d}{dr}+\frac{\kappa }{r}} G(r) \re &  -\lrs{2M+
\Sigma_{S}(r) - \Sigma_{0}(r)} F (r) + X(r)\label{HF2},
 \end{align}\esub
where the scalar potential $\Sigma_S$ and the time component of the vector
potential $\Sigma_0$ contain the contributions from the direct terms and the
rearrangement term due to the density-dependence of the meson-nucleon
couplings. {The $X $ and $Y $ functions represent the results of the non-local
Fock potentials acting on $F$ and $G$, respectively \cite{long:2006b}.}
By introducing the functions $X_G$, $X_F$, $Y_G$ and $Y_F$ as in Ref.
\cite{Bouyssy:1987},
 \bsub\label{Local}\begin{align}
X(r)  =& \frac{G(r)  X(r) }{G ^2 + F ^2} G(r)  + \frac{F(r)  X(r) }{G ^2 + F
^2} F(r) \re\equiv& X_{
G }(r) G (r) + X_{   F }(r) F(r) ,\\
Y(r)  =& \frac{G(r)  Y(r) }{G ^2 + F ^2} G(r)  + \frac{F(r)  Y(r) } {G ^2 + F
^2} F(r) \re\equiv& Y_{ G }(r) G(r)  + Y_{   F }(r) F(r),
 \end{align}\esub
the coupled differential-integral equations (\ref{RHF1}) can be transformed
into the equivalent local ones,
 \bsub\label{RHFD}\begin{align}
\lrs{\frac{d}{dr}-\frac{\kappa }{r}- Y_{   F }(r)}
F(r)  - \lrs{ \Delta(r)- E }G (r)& =0 ,\\
\lrs{\frac{d}{dr}+\frac{\kappa }{r}+ X_{   G }(r)} G(r)  +\lrs{V(r)  - E } F(r)
& =0 ,
 \end{align}\esub
where $\Delta  \equiv \Delta^D +  Y_{  G }$, $V \equiv V ^D+ X_{  F } $, and
 \bsub\begin{align}
\Delta ^D \equiv & \Sigma_S + \Sigma_0;& V ^D\equiv& \Sigma_0- \Sigma_S -2M.
 \end{align}\esub
The coupled equations (\ref{RHFD}) can be solved by using the same numerical
method as in RMF \cite{Meng:1998a}. In Eqs. (\ref{RHFD}), the functions $X_G$,
$X_F$, $Y_G$ and $Y_F$ might be taken as the effective potentials from the
exchange (Fock) terms in the DDRHF. {Eqs. (\ref{RHFD}) must be solved
iteratively since the potentials depend on the solution $(G,F)$.}

>From the radial Dirac equations (\ref{RHFD}), the Schr\"odinger-type equation
for the lower component $F $ is obtained as,
 \beq\label{RHF_F}\begin{split}
\frac{d^2}{dr^2} F  +V_1 \frac{d}{dr} F  +& \lrb{ V_\pcb + V_{\pso } + V_2}F\\
& =- \lrb{V ^D - E }\lrb{\Delta ^D - E }F ,
 \end{split}\eeq
with
 \bsub\begin{align}
V_{1 }\equiv &~ \lrb{X_{ G } - Y_{  F }} - \frac{1}{\Delta  -
E }\frac{d\Delta }{dr},\\
V_{2 }\equiv &~ Y_{  F }\frac{1}{\Delta  - E }\frac{d\Delta }{dr} - X_{ G } Y_{
F } - \frac{d}{dr}Y_{ F }\re & + Y_G \lrb{V ^D - E }
+ X_F\lrb{\Delta  - E },\\
V_{\pso }\equiv &~ \frac{\kappa }{r}\lrs{\frac{1}{\Delta  - E }\frac{d\Delta
}{dr} -
\lrb{X_{ G } + Y_{ F }}}\label{psop},\\
V_{\pcb } \equiv &~\frac{\kappa (1-\kappa )}{r^2},
 \end{align}\esub
where $V_\pcb$ and $V_\pso$ correspond to the pseudo-centrifugal barrier (PCB)
and pseudo-spin orbital potential (PSOP), respectively. In the limit of
$V_{\pso}=0$, the pseudo-spin becomes a good symmetry and the PSS can be
labeled by the pseudo radial number $\tilde n = n-1$, pseudo-orbit $\tilde l =
l'$, and pseudo-spin $\tilde s = s = \ff2$, {with the total angular momentum $j
= \tilde l \pm \tilde s$ for the two partner states.}
{For instance, the partners are $\lrb{n s_{1/2}, (n-1) d_{3/2}}$ for $\tilde l
= 1$, $\lrb{n p_{3/2}, (n-1) f_{5/2}}$ for $\tilde l = 2$,} etc. Notice that
$X_G$ and $Y_F$ in the $V_{\pso}$ are new contributions from the Fock terms
compared with RMF. The potential $V_2$ entirely originates from the Fock
contributions. The direct (Hartree) and exchange (Fock) contributions of the
PSOP and $V_{1 }$ can be separated as,
 \bsub\begin{align}
V_{\pso }^D =&~ \frac{\kappa }{r}\frac{1}{\Delta  - E }\frac{d\Delta ^D}{dr},\\
V_{\pso }^E =&~ \frac{\kappa }{r}\lrs{\frac{1}{\Delta  - E }\frac{d Y_G}{dr}
- \lrb{X_{ G } + Y_{  F }}},\label{psope}\\
V_{1 }^D =&~ -\frac{1}{\Delta  - E }\frac{d\Delta ^D}{dr},\\
V_{1 }^E =&~ \lrb{X_{ G } - Y_{  F }}-\frac{1}{\Delta  - E }\frac{dY_G}{dr},
\label{v1f}
 \end{align}\esub
while $\Delta$ {comes from} both the Hartree and Fock contributions.

To have better a understanding of the pseudo-spin orbital splitting,
especially the effects of Fock terms, it will be more transparent to
rewrite Eq. (\ref{RHF_F}) as,
 \beq\label{RHF2}\begin{split}
\frac{1}{{V ^D - E }}\frac{d^2}{dr^2} F  +\frac{1}{{V ^D - E }} &\lrs{V_\pcb +
\hat{\cals V} ^D +\hat{\cals V} ^E} F\\ &  + \Delta ^D F = E F
 \end{split}\eeq
where the operators $\hat{\cals V} ^D$ and $\hat{\cals V} ^E$ are
 \bsub\begin{align}
\hat{\cals V} ^D = &~ V_{1 }^D \frac{d}{dr} + V_{\pso }^D,\\
\hat{\cals V} ^E = &~ V_{1 }^E \frac{d}{dr} + V_{\pso }^E + V_{2
}.\label{calsVE}
 \end{align}\esub

\begin{figure}[htbp]
\includegraphics[width = 7.0cm]{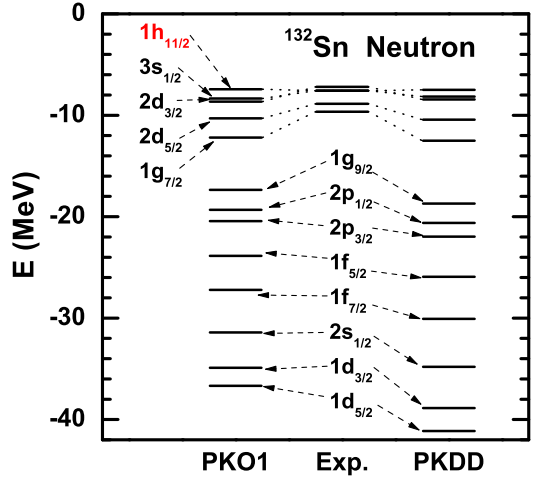}
\caption{Neutron single particle energies of $^{132}$Sn calculated by the DDRHF
with PKO1 and the RMF with PKDD. Experimental data are taken from Ref.
\cite{Oros:1996}.}\label{fig:Sn132Nlev}
 \end{figure}

Let us discuss the PSS by taking the doubly magic nucleus $^{132}$Sn
as a representative. In \figref{fig:Sn132Nlev} are shown the neutron
single-particle energies calculated by the DDRHF with PKO1, compared
with those given by the RMF with PKDD and the experimental data
\cite{Oros:1996}. Among the neutron single-particle states shown in
\figref{fig:Sn132Nlev}, there are four pseudo-spin partners $1\tilde
p$, $1\tilde d$, $1\tilde f$ and $2\tilde p$, which correspond to
the pairs $\lrb{\nu 2s_{1/2}, \nu1d_{3/2}}$, $\lrb{\nu 2p_{3/2},
\nu1f_{5/2}}$, $\lrb{\nu 2d_{5/2}, \nu1g_{7/2}}$ and $\lrb{\nu
3s_{1/2}, \nu2d_{3/2}}$, respectively. From \figref{fig:Sn132Nlev},
one can find a good PSS in the partners $\nu 3s_{1/2}$ and
$\nu2d_{3/2}$ both in the DDRHF and RMF.

\begin{figure}[htbp]
\includegraphics[width =  8.0cm]{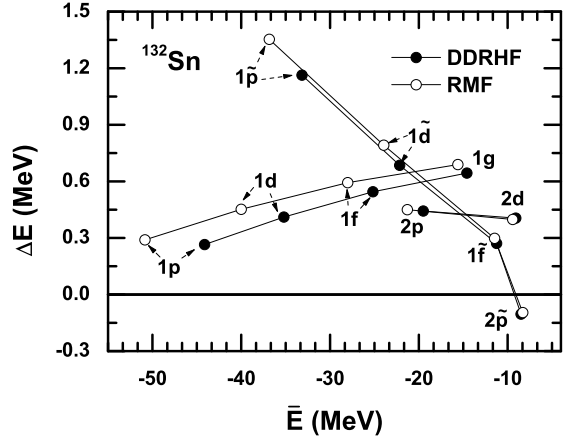}
\caption{The pseudo-spin orbital splitting $\Delta E_\pso = (E_{\tilde l j =
\tilde l - 1/2} - E_{\tilde l j = \tilde l + 1/2})/(2\tilde l + 1)$ versus the
average binding energy $\bar E_\pso=(E_{\tilde l j = \tilde l - 1/2} +
E_{\tilde l j = \tilde l + 1/2})/2$ for the neutron states in $^{132}$Sn. The
pseudo-spin partners $1\tilde p$, $1\tilde d$, $1\tilde f$ and $2\tilde p$
correspond to $\lrb{\nu 2s_{1/2}, \nu1d_{3/2}}$, $\lrb{\nu 2p_{3/2},
\nu1f_{5/2}}$, $\lrb{\nu 2d_{5/2}, \nu1g_{7/2}}$ and $\lrb{\nu 3s_{1/2},
\nu2d_{3/2}}$ states, respectively.  The spin-orbit splitting $\Delta E_\so =
(E_{l j = l - 1/2} - E_{ j = l + 1/2})/(2l + 1)$ are also given for $\lrb{\nu
1p_{3/2}, \nu1p_{1/2}}$, $\lrb{\nu 1d_{5/2}, \nu1d_{3/2}}$, $\lrb{\nu 1f_{7/2},
\nu1f_{5/2}}$, $\lrb{\nu 1g_{9/2}, \nu1g_{7/2}}$ and $\lrb{\nu 2p_{3/2},
\nu2p_{1/2}}$, $\lrb{\nu 2d_{5/2}, \nu2d_{3/2}}$ pairs as a function of $\bar
E_\so=(E_{l j + l - 1/2} + E_{ l j = l + 1/2})/2$. The results are obtained by
the DDRHF with PKO1 (filled symbols) and the RMF with PKDD (open symbols),
respectively.}\label{fig:PKO1+PKDD}
 \end{figure}

>From the single-particle energies, the pseudo-spin orbital splitting is
estimated as $\Delta E_\pso = (E_{\tilde l j = \tilde l - 1/2} - E_{\tilde l j
= \tilde l + 1/2})/(2\tilde l + 1)$ for the pseudo-spin partners $1\tilde p$,
$1\tilde d$, $1\tilde f$ and $2\tilde p$. The results are shown in
\figref{fig:PKO1+PKDD} as a function of the average binding energy $\bar
E_\pso=(E_{\tilde l j = \tilde l - 1/2} + E_{\tilde l j = \tilde l + 1/2})/2$.
For both DDRHF (filled symbols) and RMF (open symbols) results, the pseudo-spin
splitting between $\nu3s_{1/2}$ and $\nu2d_{3/2}$ is more than ten times
smaller than that between $\nu2s_{1/2}$ and $\nu1d_{3/2}$. As shown in
\figref{fig:Sn132Nlev} and \figref{fig:PKO1+PKDD}, there exist some differences
in the single-particle energies (especially the deeply bound states) between
the DDRHF and RMF, whereas the monotonous decreasing behavior of $\Delta
E_\pso$ with the decreasing binding energies is observed in the both models. As
a reference, the spin-orbit splitting $\Delta E_\so = (E_{l j = l - 1/2} - E_{l
j = l + 1/2})/(2l + 1)$ versus the average binding energy $\bar E_\so=(E_{l j +
l - 1/2} + E_{ l j = l + 1/2})/2$ is also shown in \figref{fig:PKO1+PKDD} for
the spin-orbit partners $1p$, $1d$, $1f$, and $1g$, and $2p$, $2d$. A clear
difference between the pseudo-spin orbital and spin-orbit splittings can be
seen in their energy-dependence, i.e., a strong energy-dependence is found for
the pseudo-spin orbital splitting, while the spin-orbit splitting shows a weak
energy-dependence. This difference can be understood from the comparison
between the expressions for the PSOP (\ref{psop}) and the corresponding
spin-orbit potential. The energy $E$ and the potential $\Delta$ in the
denominator $\Delta -E$ of Eq. (\ref{psop}) are comparable so that it brings a
strong energy-dependence for the PSOP. On the other hand, the corresponding
denominator of the spin-orbit potential in RHF is $V-E$ \cite{Meng:1998PRCps},
which gives much weaker energy-dependence because the single-particle energy
$E$ is much smaller than the potential $V$: the value of $E$ is a few tens MeV
while $V$ is several hundred MeV.

\begin{figure}[htbp]
\includegraphics[width = 7.0cm]{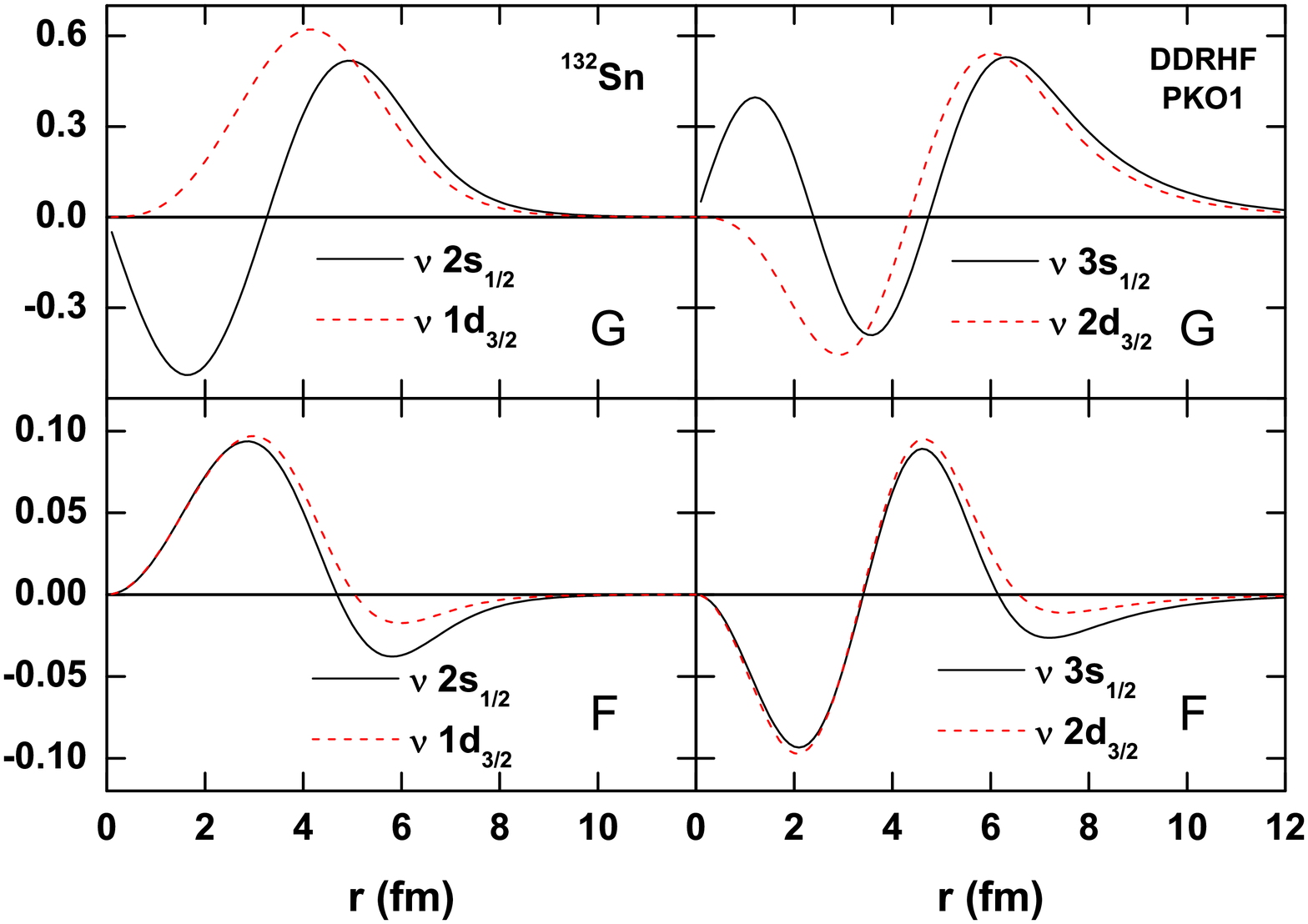}
\includegraphics[width = 7.0cm]{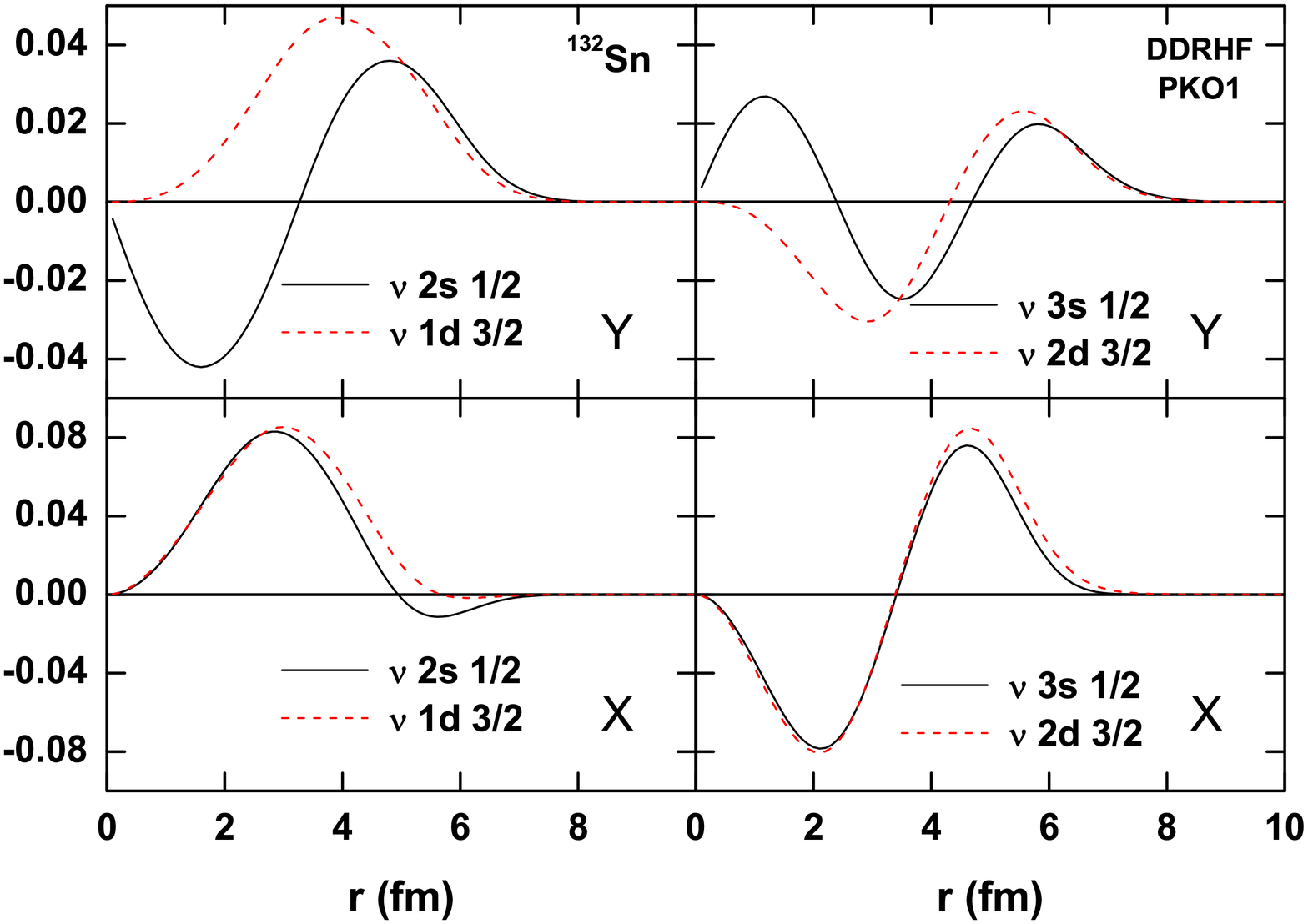}
\caption{The radial wave functions $G$ and $F$ (left panel), and the nonlocal
terms $X$ and $Y$ (right panel) in Eq. (\ref{RHF1}) given by the DDRHF with
PKO1 for the pseudo-spin partners $\lrb{\nu 2s_{1/2}, \nu1d_{3/2}}$ and
$\lrb{\nu 3s_{1/2}, \nu2d_{3/2}}$ in $^{132}$Sn. }\label{fig:NsdGFXY}
 \end{figure}

Let us discuss the effects of Fock terms, especially the {non-locality effect}
on the conservation of the PSS {by considering} the pseudo-spin partners
$1\tilde p$ and $2\tilde p$. The left panel of \figref{fig:NsdGFXY} shows the
radial wave function (upper and lower components $G$, $F$) for the pseudo-spin
partners $\lrb{\nu 2s_{1/2}, \nu1d_{3/2}}$ and $\lrb{\nu 3s_{1/2},
\nu2d_{3/2}}$. The lower components of the partner states $\nu 2s_{1/2}$
($\nu3s_{1/2}$) and $\nu1d_{3/2}$ ($\nu2d_{3/2}$) are very close each other in
both the shape and magnitude whereas the upper components are quite different.
Comparisons of the upper and lower components between the partner states
indicate that the approximate PSS exists in the lower components for both
partners $1\tilde p$ and $2\tilde p$ although the upper component of the total
Dirac wave function dominate. {The functions $X(r)$ ($Y(r)$) introduced in
Eq.(\ref{RHF1}) and representing the action of the Fock potentials on $F$ ($G$)
are also given in the right panel of \figref{fig:NsdGFXY} for the pseudo-spin
partners $1\tilde p$ and $2\tilde p$.}
It is interesting to find that the radial dependence of the {functions} $X$ and
$Y$ are very similar to those of the lower and upper components, respectively.

\begin{figure}[htbp]
\includegraphics[width = 7.0cm]{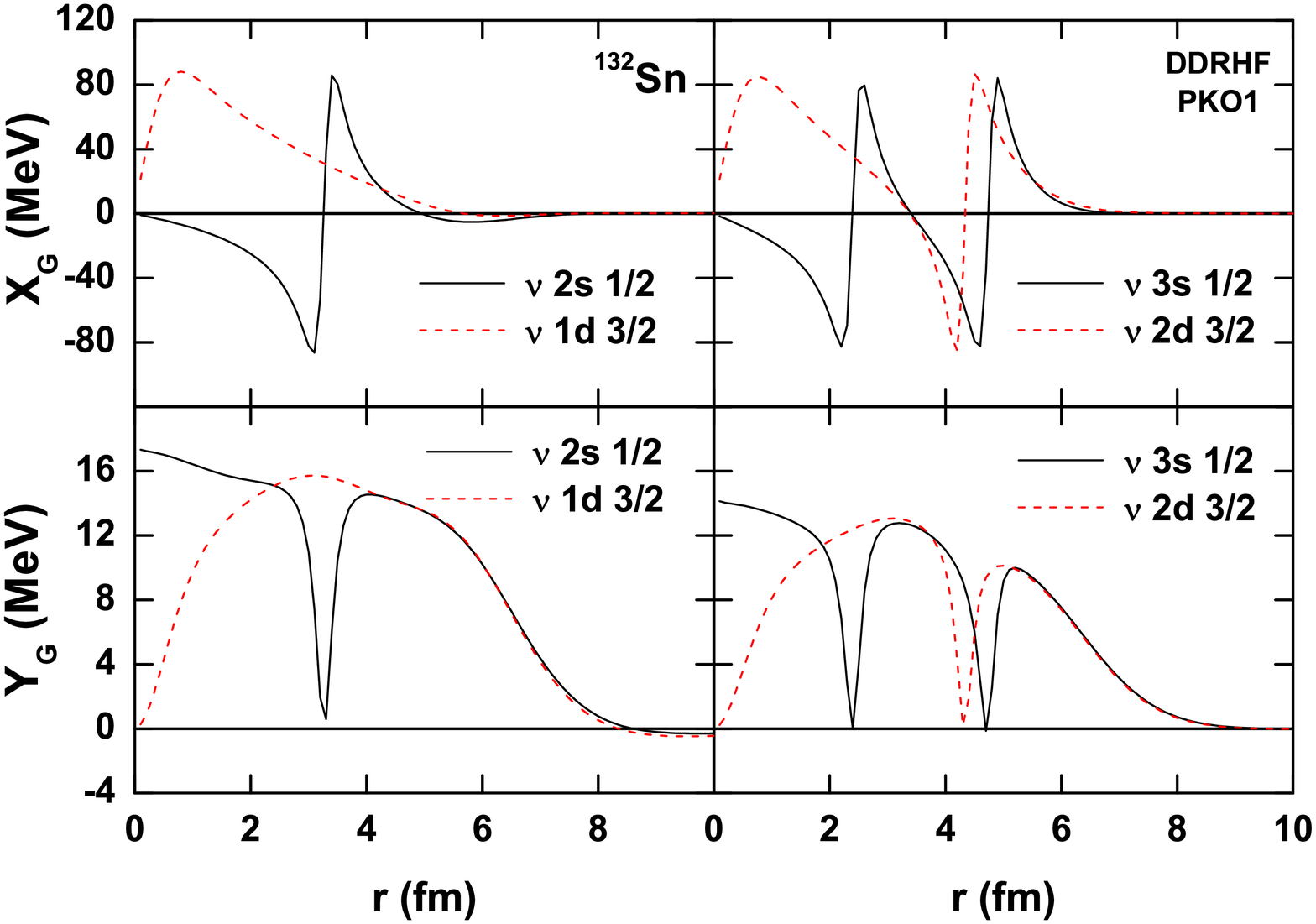}
\includegraphics[width = 7.0cm]{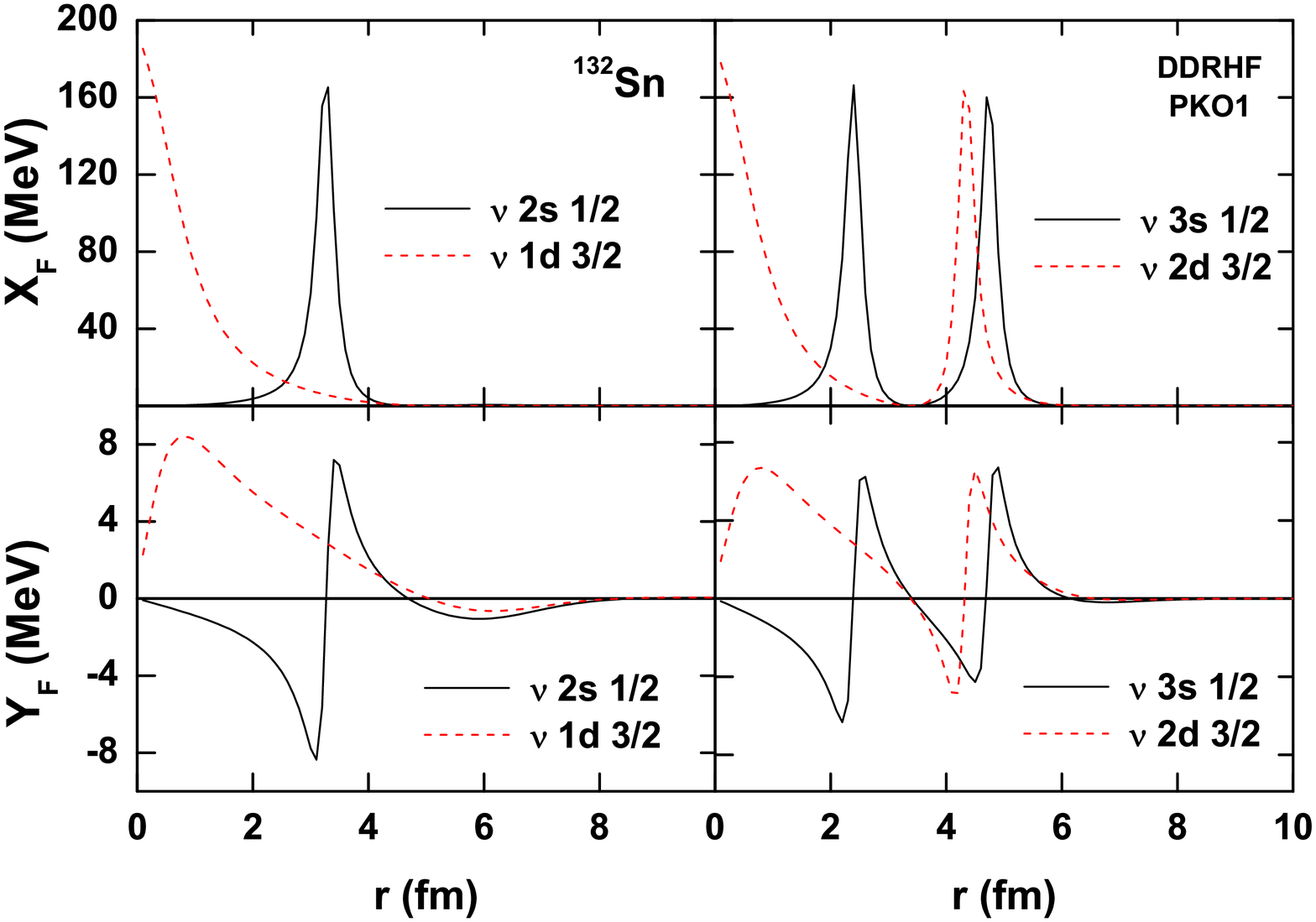}
\caption{The effective potentials $X_G$, $Y_G$ (left panel), and $X_F$, $Y_F$
(right panel) due to the exchange (Fock) terms in Eq. (\ref{RHFD}) for the
pseudo-spin partners $\lrb{\nu 2s_{1/2}, \nu1d_{3/2}}$ and $\lrb{\nu 3s_{1/2},
\nu2d_{3/2}}$ in $^{132}$Sn. See the text for detail. }\label{fig:XYGF}
 \end{figure}

In Eq. (\ref{Local}), the {Fock-related terms $X$ and $Y$}
are divided into two products of {effective potentials times} the Dirac wave
functions to obtain the equivalent local Dirac equations (\ref{RHFD}).
\figref{fig:XYGF} shows these effective potentials $\lrb{X_G, Y_G}$, and
$\lrb{X_F, Y_F}$ for the pseudo-spin partners $\lrb{\nu 2s_{1/2}, \nu1d_{3/2}}$
and $\lrb{\nu 3s_{1/2}, \nu2d_{3/2}}$. The local peaks and dips in the figures
are due to the localization of the exchange terms in Eq. (\ref{Local})
reflecting the nodes of the upper components in the denominator. It is seen in
\figref{fig:XYGF} that the radial dependence of $Y_F$ is almost identical to
that of $X_G$ since the shapes of $\lrb{X, Y}$ and $\lrb{G, F}$ are very close
each other. Because the peaks and dips in \figref{fig:XYGF} are due to the
nodes
of upper components, their contributions to the single-particle
energies will be smoothed out by the corresponding nodes. Thus, one
can ignore the non-locality represented by the peaks or dips for the
discussion on the PSS. Besides the peaks, one can find that in the
inner part of the nucleus the effects of the non-locality are
significant on $X_G$ and $Y_F$ and give a strong state dependence
with different signs, while it shows fairly weak effects on $Y_G$.
The $X_F$ is just due to the nodes of the upper component.

\begin{figure}[htbp]
\includegraphics[width =  8.0cm]{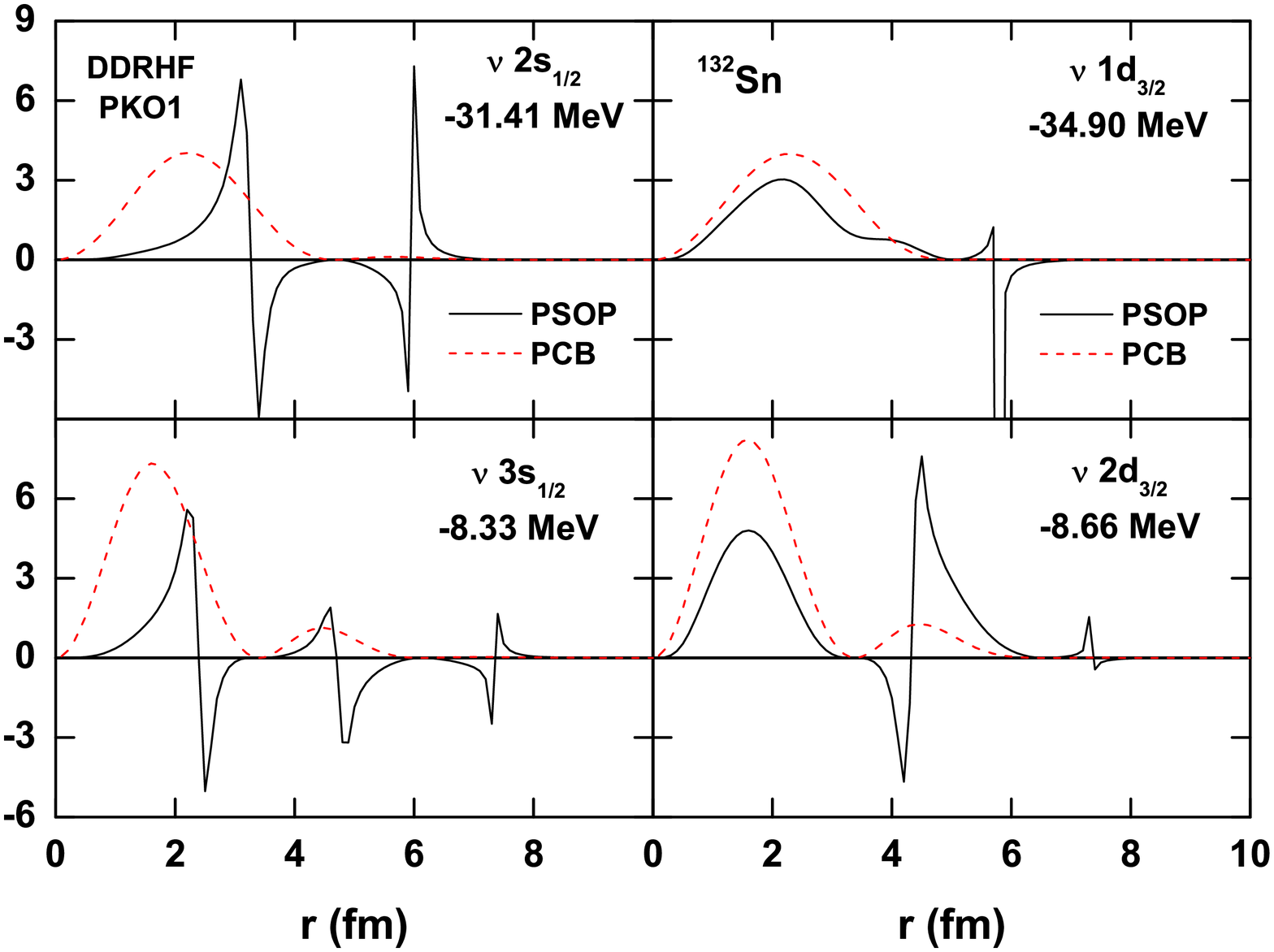}
\caption{The PCB and PSOP multiplied by the factor $F^2/\lrb{V^D- E}$ for the
pseudo-spin partners $\lrb{\nu 2s_{1/2}, \nu1d_{3/2}}$ and $\lrb{\nu 3s_{1/2},
\nu2d_{3/2}}$ in $^{132}$Sn. The PCB contributions are drawn by the dashed
 lines, while the PSOP are shown by the solid lines. The DDRHF with PKO1 is used
for the calculations. See the text for details.}\label{fig:PCB+PSO}
 \end{figure}

\begin{figure}[htbp]
\includegraphics[width =  8.0cm]{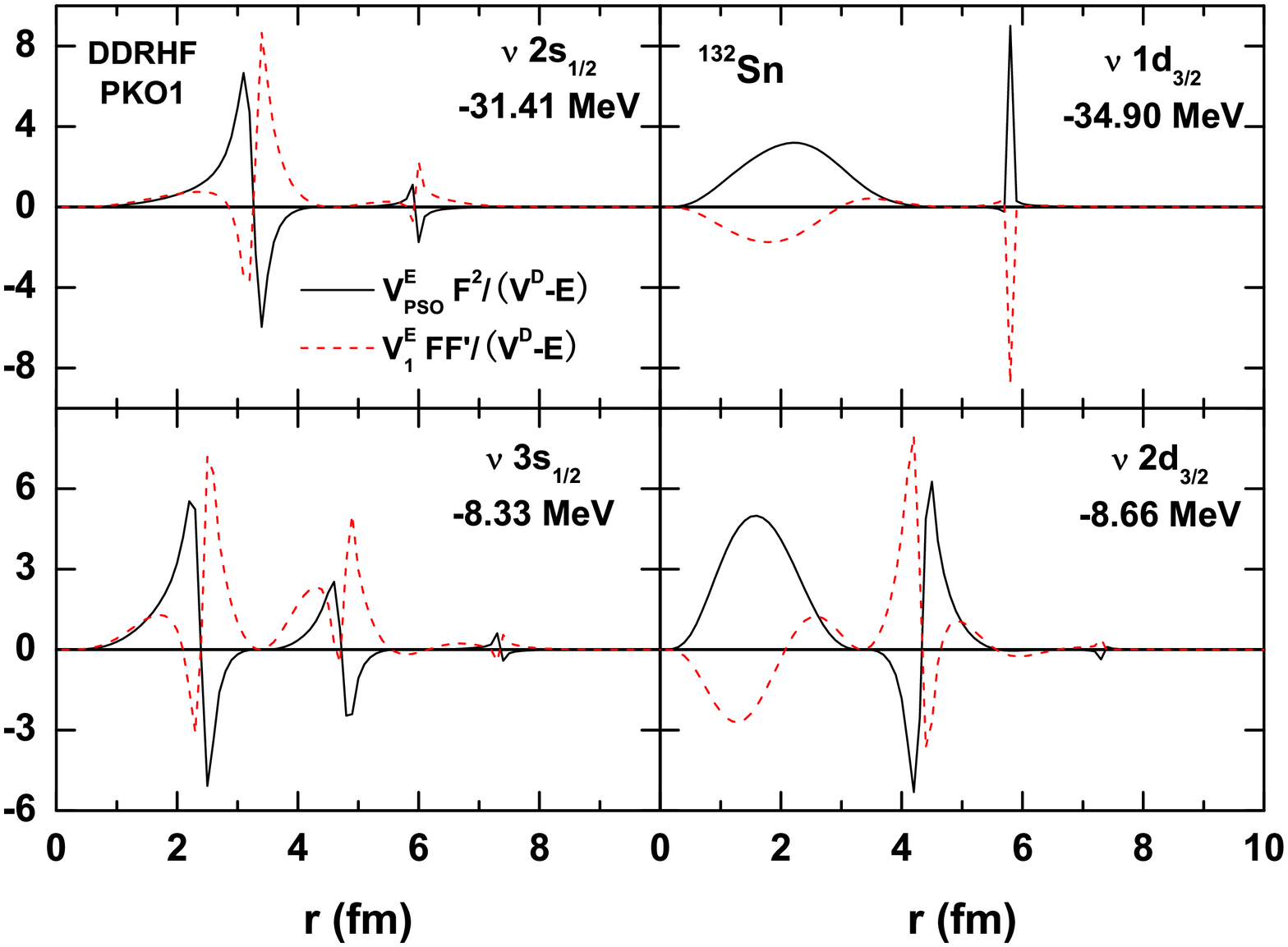}
\caption{The functions $V_{\pso}^E F^2/\lrb{V^D - E}$ and  $V_1^E
FF'/\lrb{V^D - E}$ given by the exchange (Fock) terms of the DDRHF
with PKO1 for the pseudo-spin partners $\lrb{\nu 2s_{1/2},
\nu1d_{3/2}}$ and $\lrb{\nu 3s_{1/2}, \nu2d_{3/2}}$ of $^{132}$Sn.
The singular points for $\nu 2s_{1/2}$ and $\nu1d_{3/2}$ at $r\simeq
6$ fm, and for $\nu 3s_{1/2}$ and $\nu2d_{3/2}$ at $r\simeq 7.5$ fm
are due to the denominator $\lrb{\Delta - E}$ in the PSOP (see Eq.
(\ref{psop})) while the other local peaks are due to the nodes of
the upper component $G$ (see Eq. (\ref{Local}))}\label{fig:PSOV1F}
 \end{figure}

In RMF it has been proved that the PSS
is well obeyed if the PCB is much stronger than the PSOP \cite{Meng:1998PRCps}.
\figref{fig:PCB+PSO} shows the PCB and PSOP multiplied by the factor
$F^2/\lrb{V^D- E}$ for the pseudo-spin partners $\lrb{\nu 2s_{1/2},
\nu1d_{3/2}}$ and $\lrb{\nu 3s_{1/2}, \nu2d_{3/2}}$ in $^{132}$Sn. Due to the
denominator $\Delta-E$ in Eq. (\ref{psop}), there exists singular points at
$r\simeq 6$ fm for the partner $\lrb{\nu 2s_{1/2}, \nu1d_{3/2}}$, and at
$r\simeq 7.5$ fm for $\lrb{\nu 3s_{1/2}, \nu2d_{3/2}}$. The other local peaks
in the PSOP are due to the nodes of upper component $G$. For the $s$ states
($l=0$), the PCB is much stronger than the PSOP since the contributions for the
PSOP around the nodes or the singular points are more or less {mutually
cancelling.}
On the other hand, for
the $d$ states, the PSOP are comparable to the PCB even after taking
account of the cancellation around the nodes or singular points.
Comparing the shapes of the PSOP in \figref{fig:PCB+PSO} and
$X_G,Y_F$ in \figref{fig:XYGF}, one can find that the Fock terms
present significant contributions to the PSOP, especially for the
$d$ states in the inner part of the nucleus. It is also seen that
the Fock terms in \figref{fig:PSOV1F} have substantial contributions
to the PSOP in \figref{fig:PCB+PSO}.

>From Eq. (\ref{RHF2}), one can estimate the contributions of the potentials
$V_\pcb$, $\hat{\cals V}^D$ and $\hat{\cals V}^E$ to the single-particle energy
$E$. For example, the PCB contribution can be expressed as,
 {\beq
\frac{1}{\int_0^\infty F^2 dr} \int_0^\infty \frac{V_{\pcb}}{V^D - E} F^2 dr.
 \eeq}
The results calculated by the DDRHF with PKO1 and the RMF with PKDD are shown
in \tabref{tab:SEHF} for the pseudo-spin partners $1\tilde p$ and $2\tilde p$.
For both the DDRHF and RMF, the terms $F''$, $\Delta^D$ and $\hat{\cals V}^D$
in Eq. (\ref{RHF2}) show substantial differences between the partner states
$\lrb{\nu2s_{1/2},\nu1d_{3/2}}$ and $\lrb{\nu3s_{1/2},\nu2d_{3/2}}$ whereas the
differences in the PCB and the exchange terms $\hat{\cals V}^F$ are negligible.
The differences in $F''$ and $\Delta^D$ reflect those of the lower components
{in the two} pseudo-spin partners as shown in the left panel of
\figref{fig:NsdGFXY}.


 \begin{table}[htbp]
\caption{The single-particle energies $E$ and the contributions from
different terms in the left hand side of Eq. (\ref{RHF2}) given by
the DDRHF with PKO1, in comparison with those by the RMF with PKDD.
All units are in MeV.}\label{tab:SEHF}
 \begin{tabular}{c|c|c|cc|c|c|c}\toprule[1.5pt]\toprule[0.5pt]
Model&Orbit&$E$&$F''$&$\Delta^D$&$V_{\pcb}$&$\hat{\cals V}^D$&$\hat{\cals V}^E$\\
\hline
 \multirow{4}{1.2cm}{DDRHF PKO1}
 &$\nu2s_{1/2}$&-31.41&18.11 &-75.35 & 9.30 &-2.99 &19.51\\
 &$\nu1d_{3/2}$&-34.90&14.87 &-79.01 & 9.54 & 0.44 &19.26\\\cline{2-8}
 &$\nu3s_{1/2}$& -8.33&34.25 &-72.00 &11.11 & 0.09 &18.22\\
 &$\nu2d_{3/2}$& -8.66&31.93 &-73.96 &11.32 & 3.89 &18.17\\ \hline\hline
 \multirow{4}{1.2cm}{DDRMF PKDD}
&$\nu2s _{1/2}$&-34.81&21.86 &-64.65 &11.04 &-3.07&$-$ \\
&$\nu1d _{3/2}$&-38.87&18.17 &-68.08 &11.41 &-0.37&$-$ \\ \cline{2-8}
&$\nu3s _{1/2}$& -8.15&40.13 &-61.97 &13.02 & 0.67&$-$ \\
&$\nu2d _{3/2}$& -8.44&37.65 &-63.75 &13.36 & 4.30&$-$ \\
\bottomrule[0.5pt]\bottomrule[1.5pt]
\end{tabular}
\end{table}

The large differences between the partner states can be seen in the
 $F''$, $\Delta^D$ and $\hat{\cals V}^D$ contributions. However,
these three terms cancel {largely one another}  and the PSS {is preserved to a
good degree,}
especially in the partners $\lrb{\nu3s_{1/2}, \nu2d_{3/2}}$ in both DDRHF and
DDRMF. The Fock terms $\hat{\cals V}^E$ become small although each term on the
{right hand side} of Eq. (\ref{calsVE}) shows appreciable differences between
the partner states: for example, $V_\pso^E$ is large for $d$-states. The
contributions from the exchange potentials $V_\pso^E$ and $V_1^E$ in Eq.
(\ref{RHF2}) are shown in \figref{fig:PSOV1F}. For the $s$ states, the exchange
terms $V_{\pso}^E$ {and $V_1^E$ give small contributions because of their
changing signs. On the other hand,}
for the $d$ states {there are significant cancellations between $V_{\pso}^E$
and $V_1^E$,}
especially in the inner part of the nucleus. Although the Fock terms bring
substantial contributions to the PSOP, these contributions are cancelled by the
other exchange term $V_1^E$, which stems mainly from the non-locality (the
state-dependence) of the exchange potentials. Thus, the PSS still remains
{preserved}
even after the inclusion of Fock terms due to these
large cancellations as can be seen in \tabref{tab:SEHF}.

Let us now discuss the reason why the cancellations among the
exchange terms occur. From the similar radial dependence between the
non-local terms $X$ ($Y$) and Dirac wave functions $F$ ($G$) in
\figref{fig:NsdGFXY}, we might be able to validate the following
relations,
 \begin{align}\label{Local2}
X(r) \simeq & X_0(r) F(r),& Y(r) \simeq &Y_0(r) G(r),
 \end{align}
where $X_0$ and $Y_0$ are supposed to be state-independent potentials due to
the Fock terms. Using Eq. (\ref{Local2}), the non-local RHF equations
(\ref{RHF1}) can be reduced {to local ones similar to the RMF equations in
which} the terms $X_G$ and $Y_F$ do not appear in the PSOP.
{Thus, the realization} of the PSS will be similar to the {RMF case.}
Therefore, the cancellation of the Fock contributions in \tabref{tab:SEHF} is
not occasional but it is because of the similar radial dependence between the
{Fock-related} terms $(X,Y)$ and the wave functions $(F,G)$.


In summary, the PSS in the DDRHF theory was investigated in the
doubly magic nucleus $^{132}$Sn. The PSOP was derived by
transforming the coupled radial Dirac equations into the
Schr\"odinger type equation of the lower component properly taking
account of the non-local Fock terms. The analyses of the single
particle spectrum and the pseudo-spin orbital splitting indicate
that the PSS is preserved as a good symmetry for the pseudo-spin
partner $\lrb{\nu 3s_{1/2}, \nu2d_{3/2}}$ of $^{132}$Sn in the DDRHF
on the same level as RMF, although the Fock terms bring substantial
contributions to the PSOP. These contributions to the pseudo-spin
orbital splitting, However, are cancelled by the other terms due to
the non-locality of the exchange potentials. The physical mechanism
of these cancellations was discussed in relation to the similarity
between the exchange potentials and the Dirac wave functions.

{This work is  partly supported by the National Natural Science Foundation of
China under Grant No. 10435010, and 10221003, and the Japanese Ministry of
Education, Culture, Sports, Science and Technology by Grant-in-Aid for
Scientific Research under the program number (C(2)) 16540259, and the European
Community project Asia-Europe Link in Nuclear Physics and Astrophysics
CN/Asia-Link 008(94791). }


\end{document}